%Paper: hep-ph/9306204
%From: JFGUCD@UCDHEP.UCDAVIS.EDU
%Date: Tue, 1 Jun 1993 13:52 PDT

%macropackage=phyzzx

% Date: Monday June 1, 1993
%
%   FINAL VERSION
%
\input phyzzx.tex
\input tables.tex
%\input phrdefs.tex
%\PHYSREV
%%%%%%%%%%%%%%%%%%%%%%%%%%%%%%%%%%%%%%%%%
%%%%   note that there are things behind %%% signs that must
%%%%   be brought into action for preprint version of this paper
%%%%   These include altered title page, and figure insert.
%%%%%%%%%%%%%%%%%%%%%%%%%%%%%%%%%%%%%%%%%
\overfullrule0pt
\def\prdj#1{{\it Phys. Rev.} {\bf D{#1}}}

\def\prlj#1{{\it Phys. Rev. Lett.} {\bf {#1}}}

\def\br{B}
\def\anti{\overline}
\def\to{\rta}
\def\ptlmax{E_T^{\ell\,{\rm max}}}
\def\ptjmax{E_T^{j\,{\rm max}}}
\def\mest{M_{\rm est}}
\def\gev{~{\rm GeV}}

\def\pbi{~{\rm pb}^{-1}}

\def\gevc{GeV}
\def\gevcsq{GeV}

\def\tanb{\tan\beta}

\def\gl{\widetilde g}
\def\mgl{M_{\gl}}

\def\cnone{\widetilde \chi_1^0}
\def\mcnone{M_{\widetilde \chi_1^0}}

\def\cpmone{\widetilde\chi_1^{\pm}}
\def\mcpmone{M_{\widetilde\chi_1^{\pm}}}
\def\rta{\rightarrow}

\Pubnum={LBL-34106 \cr UCD-93-17\cr  SCIPP-93/10\cr}
%\Pubnum={LBL-33422}
%\date{January 1993}
%\pubtype{T/E}
%\def\smivfoot{\footnote*{{\rm To appear in} {\it Proceedings of the 23rd
%Workshop of the INFN Eloisatron Project ``Properties of SUSY Particles''},
%{\rm Erice, September 28 - October 4, 1992}}}
%%%%%%%%%%%%%%%%%%%%%%%%%%%%%%%%%%%%%%%%%%%%%%
%%%%%% for preprint version
%%%%%%%%%%%%%%%%%%%%%%%%%%%%%%%%%%%%%%%%%%%%%%
%%%%
%%%%%%specific for Santa Barbara
%

%%%%%%
\def\wtilde{\widetilde}
\def\etal{{\it et al.}}

%Table numbering
%\def\notation{1}
%\def\cptab{2}
%\def\detectsurvey{3}
%%%%%%
%Redefinition of Gordy's equation names

%Sally's h state

%

    \def\fillbox#1#2{\hbox to #1{\vbox to #2{\vfil}\hfil}
    }
%%% general

\def\wpm{W^{\pm}}

\def\mt{m_t}

\def\tanb{\tan\beta}

\def\mz{m_Z}

\def\rta{\rightarrow}

\def\sq{\widetilde q}
\def\msq{M_{\sq}}

\def\slep{\widetilde \ell}
\def\mslep{M_{\slep}}
\def\snu{\widetilde \nu}
\def\msnu{M_{\snu}}

\def\gl{\widetilde g}
\def\chitil{\widetilde\chi}

\def\cnone{\chitil^0_1}

\def\mcnone{M_{\chitil^0_1}}

\def\wpm{W^{\pm}}

\def\LBL{\centerline{\it Physics Division}
\centerline{\it Lawrence Berkeley Laboratory, Berkeley CA 94720}}

\def\SCIPP{\centerline {\it Santa Cruz Institute
for Particle Physics}
  \centerline{\it University of California, Santa Cruz, CA 95064}}
\def\DAVIS{\centerline {\it Department of Physics}
  \centerline{\it University of California, Davis, CA 95616}}

\def\tanb{\tan\beta}

\def\chitil{\widetilde \chi}

\def\gl{\widetilde g}
\def\mgl{M_{\gl}}

\def\wpm{W^{\pm}}

\def\mz{m_{Z}}

\def\rta{\rightarrow}

\def\cnone{\chitil^0_1}

\def\mcnone{M_{\chitil^0_1}}

% some of gordy's defs along with my replacements
\def\mt{m_t}

\def\M2pm{M^2_{P^\pm}}
\def\m2z{\mz^2}
%\def\mw{M_{\ss W}}
%\def\mz{M_{\ss Z}}

%\baselineskip 12pt

%\singlespace
%\baselineskip=14pt
%\hsize=6in
%\vsize=8.5in
%\nopagenumbers

\titlepage
\noindent
%\hglue 4.8in {LBL-34106}\hfill\break
%\hglue 4.8in {January 1993}
%\vskip 0.3in

%\centerline{\bf Discovering Supersymmetry with Like-sign Dileptons\doeack}
\title{\bf DISCOVERING SUPERSYMMETRY WITH LIKE-SIGN DILEPTONS}

\author{R. Michael Barnett}
\vskip 0.10in
{\baselineskip 17pt \LBL}
\author{John F. Gunion}
\vskip 0.10in
{\baselineskip 17pt \DAVIS}
\author{Howard E. Haber}
\vskip 0.10in
{\baselineskip 17pt \SCIPP}
\sequentialequations
%\titlestyle{Like-Sign Dileptons as a Signal for Gluino Production }
%\vskip .1in
%\centerline{\twelvecp
%R. Michael Barnett$^1$, John F. Gunion$^2$, and Howard E. Haber$^3$
%}
%$^{(1)}$ {\it Lawrence Berkeley Laboratory, Berkeley, CA 94720}, and
%$^{(2)}$ {\it Department of Physics,
%University of California, Davis, CA 95616}, and
%$^{(3)}${\it Santa Cruz Institute for Particle Physics,
%University of California, Santa Cruz, CA 95064}

%%%%%%%%%%%%%%%%%%%%%%%%%%%%%%%%%%%%%%%%%%%%%%%%%%%%%%%
\vfill

%\vskip 0.3in
%\tenpoint
\centerline{\bf Abstract}

\baselineskip 20pt plus 0.2pt minus 0.1pt

Supersymmetry may be discovered at hadron colliders by searching for
events similar to the top quark signal of two isolated leptons.
In the case of gluino production, the most distinguishing feature
is that in half the events the two leading leptons have the same sign.
We demonstrate the remarkable sensitivity of this gluino signature
at both the Fermilab Tevatron Collider
and at the Superconducting Super Collider.  Techniques for
approximately determining the gluino mass are discussed.

\baselineskip 24.1pt plus 0.2pt minus 0.1pt

\vfill

\centerline{Submitted to Physical Review Letters}

\vfill
\endpage
%%%\end

\REF\gluinoref{P. Fayet, {\it Phys. Lett.} {\bf 78B}, 417 (1978);
H.E. Haber and G.L. Kane, {\it Phys. Rep.} {\bf 117}, 75 (1985).}
\REF\previousa{
R.M. Barnett, J.F. Gunion, H.E. Haber,
in {\it Proc. of 1988 Summer Study on High Energy Physics in the
1990's, Snowmass, Colorado, June 27-July 15, 1988}, edited by S. Jensen
(World Scientific, Singapore, 1989), p. 230.
}
\REF\previousb{
R.M. Barnett, J.F. Gunion, and H.E. Haber,
in {\it ``Research Directions for the Decade,'' Proc. of the 1990
Summer Study on High Energy Physics, June 25 - July 13, 1990, Snowmass, CO},
edited by E. L. Berger (World Scientific, Singapore, 1992), p. 201.
}
\REF\previousc{
H. Baer, X. Tata, and J. Woodside,
in {\it Proc. of 1988 Summer Study on High Energy Physics in the
1990's, Snowmass, Colorado, June 27-July 15, 1988}, edited by S. Jensen
(World Scientific, Singapore, 1989), p. 220.
}
\REF\previouscc{
H. Baer, X. Tata, and J. Woodside, \prdj{41} (1990) 906.
}
\REF\previousdd{
Solenoidal Detector Collaboration Technical Design Report,
E.L. Berger \etal, Report SDC-92-201, SSCL-SR-1215, 1992, p. 3-59;
GEM Technical Design Report, W.C. Lefmann \etal, Report GEM-TN-93-262,
1992, p. 2-55.
%% B.C. Barish [GEM Collaboration], invited talk at the SUSY-93
%% International Workshop, Northeastern University, March, 1993.
}
\REF\previousd{
R.M. Barnett, to appear in {\it Proceedings of the 23rd
Workshop of the INFN Eloisatron Project ``Properties of SUSY Particles'',
September 28 - October 4, 1992, Erice, Italy}, (report LBL-33422).
}
%\REF\uaone{
%C. Albajar \etal\ [UA1 Collaboration],
%{\it Phys. Lett.} {\bf B198}, 261 (1987); J. Alitti \etal\
%[UA2 Collaboration], {\it Phys. Lett.} {\bf B235}, 363 (1990).
%}
\REF\extra{G. Gamberini, {\it Z. Phys.} {\bf C30}, 605 (1986);
H. Baer, V. Barger, D. Karatas, and X. Tata, {\it Phys. Rev.}
{\bf D36}, 96 (1987).
}
\REF\bgh{R.M.~Barnett, J.F.~Gunion and H.E.~Haber, \it Phys. Rev. \bf
D37 \rm (1988) 1892.
}
\REF\rchi{R.M. Barnett, J.F. Gunion, and H.E. Haber,
\prlj{60} (1988) 401.
}
\REF\hehpdg{
H.E. Haber, {\it Phys. Rev.} {\bf D45} 1 June 1992, Part II, p.~IX.5.
}
\REF\bingun{P. Binetruy and J.F. Gunion, in INFN Eloisatron Project
Working Group Report, CCSEM Report. EL-88/1 (1988), p. 68.}

\noindent {\bf 1. Introduction}
\bigskip

A hallmark in the search for supersymmetry would be the discovery of
the gluino or the squarks, the supersymmetric partners of the gluon
and the quarks\refmark\gluinoref.
Many previous discussions of supersymmetry
phenomenology at hadron colliders have
centered on the ``classic'' signature for gluino or squark production:
clusters of hadrons (``jets'') and large missing
transverse energy with no associated hard leptons.
The missing-energy signature arises from events
in which gluinos and/or squarks are produced
and subsequently decay into jets and the lightest
supersymmetric particle (LSP).   The LSP, like the neutrino, will escape
collider detectors.
%  Note: Z\ritghtarrow\nu\anti\nu does not produce an associated lepton.
%The missing-energy signature from a prompt LSP
%differs from missing energy due to a neutrino in that no
%associated lepton is produced.

In this Letter, we focus on the striking experimental
signature of two isolated leptons which can arise
from gluino pair production. Half of the events of this type will have
leptons with the same sign of electric charge.
This signature, which is analogous to the opposite-sign lepton signal for
the top quark, may yield sensitivity much superior to the missing-energy
gluino signature at the Fermilab Tevatron Collider or at the
Superconducting Super Collider (SSC), depending on the parameters of the
supersymmetric model.  Moreover, the same-sign lepton signature can
also be utilized to provide an estimate of the gluino mass.
A preliminary version of the results contained here appeared
in Refs.~\previousa-\previousb.  Complementary work that
also considered the like-sign dilepton signature can be found in
Refs.~\previousc-\previousd.

\bigskip
\noindent {\bf 2. Gluino Decays and the Dilepton Signature}
\bigskip

In this Letter, we examine gluino searches at hadron colliders
in the context of the minimal supersymmetric
extension of the Standard Model (MSSM).
We concentrate on $\gl\,\gl$ production under the assumption
that $M_{\widetilde g}<M_{\widetilde q}$ (otherwise, $\widetilde g
\rightarrow q\widetilde q$ would be the dominant gluino decay mode,
and the phenomenology of squarks would be our primary concern).
Gluino signatures depend in detail on the
gluino branching ratios into neutralino and chargino%
\foot{The neutralinos and charginos are mixtures of the
supersymmetric partners of the $\gamma$, $Z$, and $H^0$ bosons and of the
$W^\pm$ and $H^\pm$ bosons respectively.}
final states\refmark{\extra-\rchi}.
The masses and couplings of the neutralinos
and charginos\refmark\hehpdg\ are determined by the MSSM parameters
$\mu$, $\tan\beta$ and the gaugino mass parameters,
$M_1$ and $M_2$.  Furthermore,
under a unification assumption for the gaugino mass
parameters, it follows that:
$M_2\equiv (g^2/g_s^2) \mgl$ and $M_1\equiv (5g^{\prime 2}/3g_s^2) \mgl$.

In gluino pair production, the most prominent
missing-energy signature arises from $\gl\rta q\anti q\cnone$ decays, when
$\cnone$, the lightest neutralino, is assumed to be the LSP. However,
substantial suppression of the
$\gl\rta q\anti q \cnone$ decays is found; \eg\ for $\mgl=180\gev$,
$\br(\gl\rta q\anti q\cnone)<0.4$ (and possibly much smaller, depending
upon $\mu$ and $\tanb$).
The suppression is even more dramatic for heavier gluinos
($\mgl\gsim 500\gev$), where $\br(\gl\rta q\anti q\cnone)$ is no greater
than $0.14$\refmark{\bgh,\rchi}.  The dominant gluino decays are those
into chargino and heavier neutralino states which in turn
decay into lighter neutralinos and/or charginos.
Although in a cascade decay chain the LSP is ultimately
emitted and escapes detection, the missing-energy signature is
significantly degraded as compared to gluino decays in which the LSP is
emitted in the initial decay.
Fortunately, the leptons frequently
produced in typical decay chains provide a viable signal for $\gl~\gl$
production.

Over a large range of supersymmetric parameter space,
the dominant gluino decay mode is $\gl\rta q\overline q \cpmone$,
where $\cpmone$ is the lightest chargino.\refmark\bgh\
Leptons can result from decays such as
$\cpmone\to W^{\pm}\cnone\to \ell^\pm \nu\cnone$, where the $W^{\pm}$
is either on-shell (if kinematics allow) or off-shell.  If the
sleptons are lighter than the chargino, then the two-body decays
$\cpmone\to \slep^\pm\nu \to \ell^\pm \nu\cnone$
and $\cpmone\to \widetilde\nu \ell^\pm \to \ell^\pm \nu\cnone$ are allowed
and may dominate.%
\foot{In many models of supersymmetry the $\slep$ and $\snu$
are significantly lighter than the squarks.  Thus, such decays are
not incompatible with the heavy squark assumption of this paper.}
%(we always assume here that $\msq > \mgl$). Most importantly,
Since the gluino is a Majorana fermion,%
\foot{Here, we use a broader definition of Majorana
to include neutral particles that transform under real representations
of the underlying Standard Model gauge group.}
it has the distinctive property of decaying with equal
probability into fermions and antifermions.
Thus, an excellent signature for pair production of gluinos
results from events in which both gluinos decay to a chargino of the same
sign, yielding {\it like-sign} dileptons ($\ell^+\ell^+$ or $\ell^-\ell^-$)
in the final state. The probability for the production of like-sign
and opposite-sign leptons is equal, and the characteristics
of the two classes of final states are identical.  Observation of
this distinctive result would be extremely helpful in
identifying the origin of the events.  It should also be noted
that the like-sign dilepton signature remains viable in models
with explicit $R$-parity breaking in which the LSP decays and does
not produce a missing-energy signature.%
\refmark\bingun\

Of course, like-sign dilepton events can arise in events involving primary
squark production.  If $\msq<\mgl$, dilepton signatures from $\sq\,\sq$
events could contain both like-sign and opposite-sign
components, depending on the relative squark decay branching ratios
to produce final state neutralinos and charginos.
(If $\wtilde t\rta t \cnone$ decays occur,
the additional isolated leptons from $t$ decay would need to be included
in the analysis.)
%(An especially complicated case would be $\wtilde t$ production ---
%hard isolated leptons can be emitted from the $t$ quark in
%$\wtilde t\rta t\cnone$ as well
%as from heavier neutralinos and charginos appearing in $\wtilde t$ decays.)
Since the parameters of the squark
and the $\widetilde g$/$\widetilde\chi^+$/$\widetilde\chi^0$
sectors are independent, the specific squark
branching fractions are model dependent.  As a result, we focus
in this paper on the case of $\msq>\mgl$.
It then follows that $\br(\sq\rta q\gl)\sim 100\%$ so that
$\sq\,\sq$ and $\sq\,\gl$ production yield $\gl\,\gl$ events
with two or one extra quark jet, respectively.  But,  the direct
$\gl\,\gl$ cross section is by far the largest and
yields the dominant contribution to the like-sign dilepton signal.
%Indeed, even if $\mgl$ and $\msq$ are comparable, the $\sq\,\sq$ and
%$\sq\,\gl$ cross-sections are
%significantly less than that of $\gl\,\gl$.

%%%%%!!!!!
\REF\chidecay{J.F. Gunion and H.E. Haber, \prdj{37}, 2515 (1988).}
In the numerical work in this paper, we have taken the
the branching ratio for gluino decay to the lightest chargino to be
$\br(\gl\to q\anti q^{\prime}\cpmone)= 0.58$, a result which
holds to good accuracy for all $|\mu|\gsim\mgl/3$, $m_Z$.\refmark\bgh\
Moreover, this value for the branching ratio is approximately valid for
nearly all MSSM parameters of relevance to the Tevatron gluino search.%
\foot{For example,
when $\mgl\lsim 200\gev$ and $\tanb\gsim 4$,
$\br(\gl\to q\anti q^{\prime}\cpmone)$ varies between about 45\% and 65\%
as $\mu$ is varied.}
%As noted, the Majorana nature of the gluino implies that it is equally
%likely to decay into a chargino of either sign.
If the $\slep$ and $\snu$ are heavier than the chargino,
the chargino decays dominantly
into the LSP plus a real (or virtual) $W^\pm$\refmark\chidecay,
which then decays $22\%$ of the time into electrons and muons.
Thus the branching ratio
for the decay chain $\gl \rta q\anti q \ell^\pm \nu \widetilde
\chi_1^0$ is likely to be as large as $13\%$, with equal probability
to produce a lepton of either sign.  (For simplicity,
the $\tau$--lepton will be neglected from our considerations.)
Since gluinos are produced in pairs, the
number of dilepton final states resulting from the decay of
the two gluinos would be about $1.6\%$ of all $\gl~\gl$
events, of which half would have a pair of like-sign leptons.
However, if $\mslep$ and/or $\msnu< \mcpmone$, and $\mcpmone< m_W
+\mcnone$,
then the two-body decays of the $\cpmone$ to $\slep \nu$
and/or $\snu \ell$ will have $\sim 100\%$ branching ratio into
leptonic final states (including the $\tau$-lepton).
If we neglect $\tau$--leptons, we find
$\br(\gl \rta q\anti q \ell^\pm \nu \cnone)$ close to $40\%$.  A remarkable
$15\%$ of all $\gl~\gl$ events would yield a dilepton final state.

\REF\xsections{
P.R. Harrison and C.H. Llewellyn Smith
{\it Nucl. Phys.} {\bf B213}, 223 (1983) [E: {\bf B223}, 542 (1983)];
S. Dawson, E. Eichten and C. Quigg, {\it Phys. Rev.}
{\bf D31}, 1581 (1985).}
Given the large cross-sections for
gluino pair production at the Tevatron Collider and the SSC
\refmark\xsections,
there would exist a potentially large and interesting sample of events.
These events would have hadronic
jets (two from each gluino), missing energy due to
the LSP and neutrinos in the final state, and a dilepton pair
which can come in one of the following like-sign combinations:
$e^\pm e^\pm$, $e^\pm \mu^\pm$, $\mu^\pm \mu^\pm$, and the
corresponding opposite-sign combinations.  The
events would be very similar to those arising from $t\anti t$ production
in which the leptons come from primary decays of the $t$ and $\anti t$.
Thus, distinguishing the source of opposite-sign events might be difficult.
%In fact, the $E_T(\ell)$ and isolation cuts to be
%discussed, which are designed to eliminate charm and bottom quark backgrounds,
%actually retain a larger fraction of the $t\anti t$-induced events
%than $\gl~\gl$ events. Further,
Because the efficiency for tagging $b$-jets is low and because
some fraction of $\gl\,\gl$ events will contain $b$-jets from
a hard radiated gluon, $b$-jets may not be a useful tool for
separating $t\anti t$ from $\gl\,\gl$ events.
For this reason we will focus on like-sign dilepton final states for which
$t \anti t$ production yields a background only through $\anti t\rta \anti
b \ell^-\nu$ and $t\rta b X$, $b\rta c\ell^-\nu$ (or
the corresponding charge-conjugated decay chain). This background
would be quite small since the lepton from
the $b$ decay would very rarely be isolated.

\bigskip
\noindent {\bf 3. Gluino Search at the Tevatron and SSC}

\REF\bhk{
R.M. Barnett, H.E. Haber and G.L. Kane,
{\it Nucl. Phys.} {\bf B267}, 625 (1986).}
We have evaluated the rates for the dilepton signal in gluino
pair production at
both the Tevatron and SSC energies.  In order to roughly account
for realistic experimental conditions, we have employed
a parton-level Monte Carlo, which included resolution smearing but
no fragmentation, to model the $\gl~\gl$ events\foot{A description
of a similar program to analyze the characteristics of
supersymmetric events at the CERN
$p\anti p$ collider can be found in ref.~\bhk.}
\REF\kunszt{F. Herzog and Z. Kunszt, {\it Phys. Lett.} {\bf 157B}, 430
(1985).}
We also examined the impact of the higher-order
process $\gl~\gl g$ on our results.\refmark\kunszt\
In examining the latter process, we have placed a transverse
momentum cut on the extra $g$ such that the $\gl\,\gl g$ cross section
is approximately the same as the $\gl\,\gl$ cross section.
A more precise treatment would incorporate a full one-loop
analysis, but is unlikely to affect our basic conclusions.
We find that the only impact of the extra radiated gluon is
upon the size of the cross section even when cuts
are imposed. The shapes of the distributions
we have examined (described below) are essentially unaltered
at both Tevatron and SSC energies.

We first present our analysis for Tevatron energies.
The surprisingly large potential for gluino discovery at the Tevatron
becomes apparent by giving the number of dilepton
(opposite- plus like-sign) events obtained in a 25$\pbi$ year.
For the case where the $\cpmone$
decays to $\slep\nu$ or $\ell\snu$, the net branching
ratio of 15\% quoted above yields roughly 1440, 476, 183, 79, 37, and 18
dilepton events (before cuts) for gluino masses of 100, 120, 140, 160, 180, and
200 GeV, respectively. Dilepton rates originating from
$\cpmone\to W^{(*)} \cnone$ decays are a factor of roughly 9 smaller.

\REF\leplimit{
A complete summary of results from all LEP collaborations
can be found in K. Hikasa \etal\ [Particle Data Group], {\it Phys. Rev.}
{\bf D45} 1 June 1992, Part II, pp.~IX.5--IX.12.}
\REF\gluinolimit{F. Abe \etal\ [CDF Collaboration], {\it Phys. Rev.
Lett.} {\bf 69}, 3439 (1992).}
To estimate the rates for {\it detectable} dilepton
events at the Tevatron, we assume a trigger which requires that
the leading (secondary) lepton has $E_T> 15$ (10) GeV.
In addition, we require $|\eta|<2.5$ and isolation for both
leptons.\foot{In this paper, we define an isolated lepton to be one
that is separated by at least 0.3 units in $\Delta R\equiv
[(\Delta\eta)^2+(\Delta\phi)^2]^{1/2}$ from any parton (or ``merged
parton jet'' if two or more partons are within 0.7 units in
$\Delta R$ of each other).}
Associated hadronic jets are not required.
The probability that events pass these cuts depends in
detail upon the masses of the particles involved in the decay
chains, but is roughly 50\% for much of parameter space.
We have explored a range of supersymmetric parameters
for which the mass $\mcpmone$ varies
between about 45~GeV (its current lower bound from
LEP data\refmark\leplimit) and 80~GeV,  while letting the
gluino mass vary between 100 GeV (the present experimental lower
bound\refmark\gluinolimit) and 200 GeV. In the corresponding
range of MSSM parameter space, the mass
of the LSP ($\cnone$) is approximately given by $\mgl/6$.

For $\cpmone\rta {\wpm}^*\cnone$, the 1.6\% net branching ratio
quoted earlier yields between 70 and 1 dilepton events
(of which half are like-sign) per 25 pb$^{-1}$ at the Tevatron
for $\mgl$ between 100 and 160 GeV.   In contrast, if the $\cpmone$ decays to
$\slep\nu$ and/or $\ell\snu$, the larger 15\% net branching ratio can
result in up to 1100 (15) dilepton events for $\mgl=100$ (200) GeV,
depending on the decay mode and the various masses.

\TABLE\tevatron{}
\topinsert
\noindent{\twelvepoint
\baselineskip 0pt
Table~\tevatron: Number of $\ell^\pm\ell^\pm$ plus $\ell^\pm\ell^\mp$
events after lepton cuts for various $\mgl$ values at the Tevatron with
$\int {\cal L}=25\pbi$. The various decay modes of
the $\cpmone$ are indicated by $W$ ($W\cnone$), $\slep$ ($\slep \nu$),
$\snu$ ($\snu\ell$). These rates assume the branching ratios
quoted earlier (which are, in fact, typical over a wide range of
parameters).}
\bigskip

\thicksize=0pt
\hrule \vskip .04in \hrule
\begintable
Mode | $W$ & $W$ & $W$ & $\slep$ & $\snu$  & $\snu$  & $\slep$  & $\slep$
 & $\slep$ &  $\snu$ &  $\snu$ &  $\snu$ &  $\slep$ &  $\snu$ \cr
$M(\gl)$ | 140 & 140 & 140 & 160 & 160 & 160 & 180 & 180 & 180 & 180 & 180 &
 180 & 200 & 200 \cr
$M(\cpmone)$ | 80 & 60 & 45 & 60 & 80 & 60 & 80 & 80 & 45 & 80 & 80 &
 60 & 80 & 80 \cr
$M(\slep~{\rm or}~\snu)$ | $-$ & $-$ & $-$ & 40 & 70 & 50 & 75 & 55 &
 40 & 40 & 60 & 50 & 75 & 40 \cr
 Events | 9 & 6 & 3 & 22 & 9 & 12 & 30 & 22 &
9 & 28 & 20 & 7 & 15 & 15 \endtable
\hrule \vskip .04in \hrule
\endinsert

To illustrate, we present in Table~\tevatron\ $L=25\pbi$
dilepton event rates for a sampling of cases which yield $\lsim 30$ events.
For comparison, with our cuts, 12 events are expected
from $t\anti t$ production for $\mt=140\gev$, but all of these have
opposite-sign leptons. Thus,
for comparable gluino and top quark masses, the additional requirement of
two isolated {\it like-sign} leptons  would reduce the top quark rate
to a level far below that from gluinos.  However, a top quark significantly
lighter than the gluino might require that additional cuts be made to separate
the two signals.

The event rate depends strongly on the lepton cuts. The lepton spectrum
itself is quite sensitive to decay modes and decay product masses.
For the chargino three-body decay to $\cnone\ell\nu$ via a virtual $W$,
the lepton spectrum depends primarily on $\mcpmone-\mcnone$.
For chargino decays to $\slep \nu$ ($\snu\ell$),
the spectrum is essentially determined by
$\mslep-\mcnone$ ($\mcpmone-\msnu$).  Because of
the current experimental limit of
$\mslep>44$ GeV\refmark\leplimit,  most leptons from $\slep$ decay
pass our cuts for the $\mcnone$ values employed.
Thus, even for the relatively large gluino masses of 160 and 180 GeV,
the $\slep\nu$ decay event rates illustrated
are large enough to be in possible conflict with observed
rates at the Tevatron. In contrast, as illustrated in Table~\tevatron,
the event rate associated with $\cpmone\rta \ell\snu$
decays could be very small
since small values of $\mcpmone-\msnu$ are possible, leading to a soft
lepton that is unlikely to pass our cuts.
However, even for $\mgl=180\gev$, if the $\snu$
mass is not close to $\mcpmone$ then the event rate
for the $\ell\snu$ mode is large.

Thus, if very few or no like-sign
dilepton events are found after accumulating
$L=25\pbi$, then improved limits on the gluino mass (as a function of other
MSSM parameters) will be attainable.  If $\cpmone\rta {\wpm}^*\cnone$ is the
dominant decay, a modest improvement of $\mgl>120$~GeV is possible, based
on the like-sign dilepton search.
In contrast, if $\ell\snu$ or $\slep\nu$ decays of the $\cpmone$ are dominant
and the $\ell$ spectra are not suppressed by a small mass difference,
then  limits of order $\mgl\gsim 200$ GeV will be obtained
over the large region of parameter space for which
$\br(\gl\rta q\anti q \cpmone)$ is substantial.

Assuming that one has succeeded in isolating gluino candidates, it
is important to ask if one can estimate the
mass of the gluino, and the masses
of the decay products.  We have performed detailed studies and find that
the most useful distributions (at the Tevatron) for this purpose
are $\ptlmax$, the $E_T$ of the most energetic lepton and $\ptjmax$, the
transverse energy of the jet with largest $E_T$.
\foot{The $E_T$ spectrum of the jets is completely determined by
$\mgl-\mcpmone$.} For any given decay chain,
these can be used to estimate the mass differences, and the overall event rate
can then be used to determine the absolute mass scale, provided
statistics are adequate.  Scenarios consistent with
current Tevatron rates would require $L> 100\pbi$ to achieve a
$\pm25\gev$ uncertainty in the measurement
of $\mgl$. Further details will be given elsewhere.

We now turn to the gluino search via the dilepton signature at the
SSC.  We consider only the case of $\mslep$, $\msnu > \mcpmone$
(where event rates are lowest).
%Except as noted,
%we used the GUT predictions for $M_\cpmone$ and $M_\cnone$ for large
%values of the $\mu$ parameter in the MSSM.
Before applying cuts, a 180 \gevcsq\ (2 TeV)
gluino will result in roughly $4\times 10^{6}$ (50) dilepton
events (half of which are like-sign)
in an SSC year ($10^4~{\rm pb}^{-1}$).  We triggered
on leptons with large $E_T$ in the $|\eta|<2.5$ region;
either both leptons must have $E_T>20$ \gevc, or
one lepton must have $E_T>15$ \gevc\ and the other $E_T>40$ \gevc.
In the following analysis only events for which
there are at least two jets in the region $|\eta|<3$, having
$E_T>25$ GeV are accepted.
A circularity cut of $C<0.6$ was imposed,
where circularity is defined by
\hbox{$C={1\over 2} {\rm min}\,
(\Sigma \vec E_t \cdot \hat n)^2/(\Sigma E_t^2)$}.
Here, the sum is taken over calorimeter cells and the minimization is
performed over all unit vectors, $\hat n$,
in the transverse plane. $C=0$ implies pencil-like
events and $C=1$ corresponds to isotropic events.

To determine the gluino mass, the event rate must be used in addition to
the $\ptlmax$ and $\ptjmax$ distributions which only determine mass
differences in the decay chain.
To give an example, in the $W$-mediated decays
the distributions for the cases
$[\mgl,\mcpmone,\mcnone]=[300,100,50]$ GeV and [350,150,100] GeV
are intrinsically indistinguishable.
However, the event rates after our basic cuts are very different:
assuming the effective branching ratio of 1.6\% discussed previously,
we find 87,000 versus 44,000 like-sign events per SSC year.  In
this range of gluino masses there is a sufficient number of events
and a 15\% determination of $\mgl$ would be possible
(if theoretical and experimental normalizations are reliable).
%after fixing relative mass values as a function of
%$\mgl$ using the $\ptlmax$ and $\ptjmax$ shapes.
It should be noted that we are using the fact that
$\br(\gl\rta q\anti q\cpmone)$ is almost always near its asymptotic value
for $\mgl$ masses in the range considered.
Of course, if the mass of the $\cpmone$ is known from other data
(\eg\ LEP-II), the $\ptjmax$ distribution shape
in principle allows a fairly accurate
$\mgl$ determination for a given decay scenario, independent of
the overall event rate.
%Statistics at the SSC will be sufficient
%such that this can be done in practice.

Although $\ptlmax$ and $\ptjmax$ and event rate are the most fundamental
indicators of mass differences and the overall mass scale,
an effective gluino mass variable, $\mest$, is also very useful. For
the subset of events having four jets with $E_T>100,60,60,50\gev$ and
$C<0.3$, we have computed the combined mass of all leptons and jets in
the two circularity hemispheres; $\mest$ is defined to be the larger
of these two masses. $\mest$ is very sensitive to changes in $\mgl$ for
fixed ratios of decay product masses to $\mgl$.

\FIG\figone{For events with two isolated like-sign leptons, we
illustrate the distribution of $\mest$, a variable defined in the
text that is sensitive to the gluino mass.  The curves shown
correspond to one year of SSC running ($10^4$~pb$^{-1}$), with
gluino masses of $\mgl=300$ (solid) and $350\gev$ (dots).
Dominance of the $\cpmone\rta W^* \cnone$ decays with
$\mcpmone$ and $\mcnone$ is assumed.
The distributions have been scaled so as to allow comparison
of the shapes.}
\topinsert
%\vskip 2.65in
\vbox{\phantom{0}\vskip 4in
\phantom{0}
\vskip .5in
\hskip +0pt
\special{ insert scr:short.ps}
\vskip -1.4in }
\centerline{\vbox{\hsize=12.4cm
\Tenpoint
\baselineskip=12pt%\parindent=1pc
\noindent
Figure~\figone:
For events with two isolated like-sign leptons, we
illustrate the distribution of $\mest$, a variable defined in the
text that is sensitive to the gluino mass.  The curves shown
correspond to one year of SSC running ($10^4$~pb$^{-1}$), with
gluino masses of $\mgl=300$ (solid) and $350\gev$ (dots).
Dominance of the $\cpmone\rta W^* \cnone$ decays with
$\mcpmone$ and $\mcnone$ is assumed.
}}
\endinsert
%%%

The value of the variable $\mest$ can be illustrated using $W$-mediated
decays
of the $\cpmone$ in the limit of $|\mu|\gg M_1$, $M_2$ where
$\mcnone\sim\mgl/6$ and $\mcpmone\sim\mgl/3$.\refmark\chidecay\
In Fig.~\figone\ we give the $\mest$
distribution shapes that would be obtained after one SSC year
for $\mgl=300$ and $350\gev$.
The distributions contain 4500 and 4000 events, respectively.
We see that $\mest$ provides a clear separation between the
two gluino mass cases considered.  We estimate that a
gluino mass determination accurate to within 25 GeV should be
possible in the mass range considered here.  In contrast,
the $\ptlmax$ and $\ptjmax$ histograms (not shown) provide a much weaker
gluino mass determination, barely allowing one
to distinguish $\mgl=350\gev$ from $\mgl=300\gev$.
%%Indeed, by combining the lepton and jet information
%%with jet cuts that focus on the high-$E_T$ tails, $\mest$ provides
%%a very sensitive measure of differences that are much less
%%apparent in the $\ptlmax$ and $\ptjmax$ distributions.
However, using the $\mest$ distribution, we estimate that
the gluino mass can be determined to better than
10\% if the gluino decay product masses are known.
This determination can
be cross-checked with theoretical expectations for the total event rate.
%%unless there are large systematic uncertainties in predictions for the
%%latter.

\bigskip
\noindent {\bf 4. Conclusions}
\bigskip

The like-sign signature for $\gl\,\gl$ production provides a
powerful tool, both for discovering evidence for supersymmetry and
for estimating the gluino mass.
It is the Majorana nature of the gluino that yields this
striking signature.  At the Tevatron collider,
%(with present published mass limits)
under the most favorable assumptions concerning gluino cascade decay branching
fractions, gluino production could yield
as many as 1100 dilepton events (after significant lepton cuts)
in the current 25 pb$^{-1}$ run, of which half would be like-sign.
Since large numbers of events are not seen, many new constraints on the
masses of the $\gl$, $\cpmone$, $\cnone$, $\slep$ and $\snu$ will be
obtained. If the handful of dilepton events currently observed at the
Tevatron were due to $\gl~\gl$ production, an integrated luminosity
of at least $L=100\pbi$ would be required to obtain a $\pm15\%$ estimate
of the gluino mass.
At the SSC, very large like-sign dilepton event rates are predicted, and a
gluino mass determination would be possible for masses up to 2 TeV or more.
For $\mgl<1$~TeV, the gluino mass determination would be accurate to
about 15\% (or even better if the gluino decay product masses are known).
For a very heavy gluino, like-sign dilepton events
provide the cleanest discovery signature.

%%%%%!!!!!

\vskip .1in
\centerline{\bf Acknowledgements}
\vskip .1in

This work was supported in part by the Director, Office of
High Energy and Nuclear Physics, Division of High Energy Physics,
of the U.S. Department of Energy under Contract Nos.
DE-FG-03-91ER40674, DE-AC03-76SF00098,
DE-AS03-76ER70191, and DE-AA03-76-SF00010, and by the National
Science Foundation under agreement no. PHY86-15529.

\refout
%\figout

\bye